\documentstyle[twocolumn,aps]{revtex}
\begin{document}
\draft

\title{Comment on ``Limits of the measurability of the local quantum
electromagnetic-field amplitude"}
\author{V. Hnizdo}
\address
{Department of Physics, Schonland Research Centre for Nuclear Sciences,
and Centre for Nonlinear Studies, \\
University of the Witwatersrand, Johannesburg, 2050 South Africa
\newline \newline
({\rm Present address}: National Institute for Occupational 
Safety and Health, 1095 Willowdale Road, Morgantown, WV 26505)
}
\maketitle
\begin{abstract}
It is argued that
the findings of a recent reanalysis by Compagno and Persico
[Phys. Rev. A {\bf 57}, 1595 (1998)] of the
Bohr--Rosenfeld procedure for the measurement of a single space-time-averaged
component of the electromagnetic field are 
incorrect when the field measurement time is shorter than that
required for light to traverse the measurement's test body.
To this end, the  time-averaged ``self-force" on the test body, assumed for 
simplicity to be of a spherical shape,
is evaluated in terms of a one-dimensional quadrature for the
general trajectory allowed for the test body by Compagno and Persico, and in
closed form for the limiting steplike trajectory used by Bohr and Rosenfeld.
\end{abstract}

\pacs{PACS number(s): 12.20.Ds}

In a recent paper, Compagno and Persico (CP)  \cite{CP}
revisited the famous analysis of the measurability of the electromagnetic
field by Bohr and Rosenfeld (BR) \cite{BR}.
CP analyze the BR procedure for the measurement of a single
space-time-averaged component of the electromagnetic field by treating
the interaction of an extended test body with the local quantized
electromagnetic field quantum-mechanically in the electric dipole
approximation, which is valid for field measurement times $\tau >a/c$
($a$ characterizes the linear dimensions of the test body). They obtain
a minimum uncertainty in the measured field component that they claim is
different from that obtained by BR and which the latter authors eliminated 
by connecting the test body to the reference frame by a compensating spring.
CP eliminate their minimum uncertainty simply by removing
from the measurement the neutralizing body employed in the BR
procedure to minimize the test body's field effects.

To investigate why their findings differ from the widely accepted
BR results, CP re-calculate the force on the test body due to the field
created by the neutralizing body and the test body itself
using classical electrodynamics as the BR analysis,
and thus not restricting the field measurement times $\tau$
to only $\tau >a/c$. However, CP here relax the BR assumption
that the test body's unpredictable displacement resulting from the initial
momentum measurement stays constant throughout the time period of the field
measurement. Using this calculation, CP confirm the results
of their quantum-mechanical treatment of the problem, and identify
the reason for what they believe is the difference between their results
and those of BR
in the approximation, according to CP incorrect, of a constant displacement
of the test body, which allows us to take the test body's trajectory outside
a time integration and recovers the BR results.
CP thus draw a far-reaching conclusion that a single
space-time-averaged
component of the electromagnetic field can be measured with arbitrary
accuracy without any use of compensating forces even when the field
measurement time $\tau <a/c$; in their opinion, the
necessity for compensating forces of non-electromagnetic nature would
indicate that
quantum electrodynamics is not self-consistent as a physical theory.

Using the Fourier-transform methods of a recent work \cite{VH}, where
the geometric factors of the field commutators and spring
constants employed in the BR analysis are calculated,
we evaluate here, as explicitly as is possible in general terms,
the time-averaged force on a spherical test body that is due to the
fields, calculated assuming classical electrodynamics, of both the test and
neutralization bodies; following CP, we call this force the
test body's average self-force. Using this evaluation,
we show that the limiting average self-force obtained by BR with a steplike
trajectory of the test body's constant displacement
approximates well the time-averaged self-force obtained with
a trajectory that, while conforming to the condition that the test body's
maximum speed $v_{\rm max}\ll c$, approaches sufficiently closely
the BR steplike trajectory.
This provides a rigorous justification of the fact that the use of a
steplike trajectory is fully consistent with the physical assumptions of
the BR analysis, and refutes the implication of CP that such
an approximation is incorrect.

We show also that the BR average self-force
for a given field measurement time $\tau<a/c$
has a component that is the steplike-trajectory limit of the time average of
what CP call the radiation-reaction component of
the self-force and which is not affected by the removal of the neutralizing
body. Contrary to the conclusion of CP,
this implies the need for a BR  compensating spring even if the removal
of the neutralizing body would leave
the ``radiation-reaction" component as the net self-force. This is because
the time average of the ``radiation-reaction" component for
field measurement times $\tau<a/c$ cannot be reduced
arbitrarily when the test body's trajectory is of a sufficiently steplike
character---and such a kind of trajectory is necessitated by the
requirements on the type of momentum measurements that have to be performed
on the test body.

Our starting point is expression (37) of CP, which they obtained for
the self-force on a test body that describes a trajectory $Q(t_1)$ along a
given direction, say the $x$-direction, during the field measurement period
$0\le t _1\le\tau$.
This self-force is due to the fields of the test body itself and of a
neutralizing body charged oppositely and
occupying permanently the space region of the test body's initial location,
and CP assumed in its derivation that
the test body's displacement $Q$ and velocity $\dot{Q}$ are such that
$|Q|\ll a$ and $|\dot{Q}|\ll c$.
We assume that the test body has a constant charge density
$\rho_c$ and is spherical with radius $R$, and describe its spatial region
using a uniform distribution normalized to unit volume,
$\rho(r)=(1/V)\Theta(R-r)$,  $V=(4/3)\pi R^3$.
The test body's self-force $F(t_2)$ for $0\le t _2\le\tau$ is thus given as
\begin{eqnarray}
F(t_2)&=&\rho_c^2V^2\int\rho(r_1)\,d{\bf r}_1
\int\rho(r_2)\,d{\bf r}_2\nonumber \\
&&\times\int_0^{\tau}dt_1 Q(t_1)A^{(1,2)}_{xx}(t,{\bf r}),
\end{eqnarray}
where the quantity  $A^{(1,2)}_{xx}$ is the distribution
\begin{equation}
A^{(1,2)}_{xx}(t,{\bf r})=-\left(\frac{\partial^2}{\partial x_1\partial x_2}
-\frac{\partial^2}{\partial t_1\partial t_2} \right)
\frac{\delta(t-r)}{r},
\end{equation}
with $t=t_2-t_1$, ${\bf r}={\bf r}_2-{\bf r }_1$, and $r=|{\bf r}|$.
Units in which the speed of light $c=1$ are used henceforth.

The quantity of our interest is the time-averaged self-force $\bar{F}=
(1/\tau)\int_0^{\tau}dt_2F(t_2)$, which can be written using Eq.\ (1) as
\begin{equation}
\bar{F}=\frac{\rho_c^2V^2}{\tau}\int_0^{\tau}dt_1Q(t_1)f(t_1),
\end{equation}
where
\begin{equation}
f(t_1)=\int\rho(r_1)\,d{\bf r}_1\int\rho(r_2)\,d{\bf r}_2\int_0^{\tau}
dt_2 A_{xx}^{(1,2)}(t,{\bf r}).
\end{equation}
With the spherically symmetric distribution $\rho(r)$, only the
monopole component $A^{(1,2)}_{xx\,0}(t, r)$ in a multipole expansion of
the distribution $A^{(1,2)}_{xx}(t,{\bf r})$
contributes to the double space integral in (4),
\begin{equation}
A^{(1,2)}_{xx\,0}(t, r)=-\frac{2}{3}\frac{\delta''(t-r)}{r}
-\frac{2}{3}\lim_{\epsilon\rightarrow 0}
\frac{\epsilon}{(r+\epsilon)^3}\frac{\delta(t-r)}{r}.
\end{equation}
Here, the second term arises from a regularization of the space derivative
part in (2) by
\begin{equation}
\frac{\partial^2}{\partial x_1\partial x_2}\frac{\delta(t-r)}{r}
=\lim_{\epsilon\rightarrow 0}
\frac{\partial^2}{\partial x_1\partial x_2}\frac{\delta(t-r)}{r+\epsilon},
\end{equation}
where the limit $\epsilon\rightarrow 0$ is understood to be taken only after
a two-dimensional integration;
it is the only term with which the regularization can  contribute
to the multidimensional integral (4) that defines the function $f(t_1)$.
To evaluate this integral, we perform first the integration of the 
monopole component (5) with respect to time $t_2$:
\begin{eqnarray}
\bar{A}^{(1,2)}_{xx\,0}(t_1, r)&=&\int_0^{\tau}dt_2\,A^{(1,2)}_{xx\,0}(t, r)
=-\frac{2}{3}\frac{\delta'(\tau-t_1-r)}{r}\nonumber \\
&&-\frac{2}{3}\lim_{\epsilon\rightarrow 0}
\frac{\epsilon}{(r+\epsilon)^3}\frac{\Theta(\tau-t_1-r)}{r}.
\end{eqnarray}
Here, use was made of the fact that $\delta'(-t_1-r)=0$ and
$\Theta(t_1+r)=1$ for $t_1>0$.
The function $f(t_1)$ of Eq.\ (4) is now given by 
\begin{equation}
f(t_1)=\int\rho(r_1)\,d{\bf r}_1\int\rho(r_2)\,d{\bf r}_2
\bar{A}^{(1,2)}_{xx\,0}(t_1, r),
\end{equation}
where the double space integration can be done in closed form using
the Fourier-transform method for evaluation of folding integrals \cite{VH}:
\begin{eqnarray}
f(t_1)&=&-\frac{6}{\pi R^2}\int_0^{\infty}[j_1(qR)]^2\{2\cos[q(\tau-t_1)]+1\}
\,dq
\nonumber \\
&=&\frac{1}{2R^3}(\chi-2)(2-2\chi-\chi^2)\Theta(2-\chi)-\frac{1}{R^3},
\end{eqnarray}
where $\chi=(\tau-t_1)/R$.
The above momentum-space integral involves
the Fourier transform $3j_1(qR)/qR$ of the uniform distribution
$\rho(r)$, and the Fourier transform
$-(4\pi/3)\{2\cos[q(\tau-t_1)]+1\}$ of the ``folding" function
$\bar{A}^{(1,2)}_{xx\,0}(t_1, r)$; its evaluation was done with
the help of the computing system {\it Mathematica} \cite{Wolf}.
Equations (3) and (9) give the average self-force $\bar{F}$ in terms of a
one-dimensional
quadrature involving the test-body's so-far unspecified  trajectory $Q(t_1)$.
We now assume that the trajectory $Q(t_1)$ is of a steplike character, i.e.,
in an initial time interval $(0,\Delta t)$ with $\Delta t\ll\tau$,
the displacement $Q(t_1)$ goes smoothly from 0 to a value $Q$, then
$Q(t_1)=Q={\rm const}$ for $\Delta t\le t_1\le \tau-\Delta t$,
and in a final interval $(\tau-\Delta t,\tau)$ the displacement $Q(t_1)$
returns smoothly from $Q$ back to 0. As the test body's maximum speed
$v_{\rm max}\equiv{\rm max}|\dot{Q}(t_1)|$, $0\le t_1\le \tau$ 
must satisfy the condition $v_{\rm max}\ll c$, the  constant $Q$ is such that
$|Q|<v_{\rm max} \Delta t\ll c\Delta t$, and if one defines the mean speed
$\bar{v}$ in the initial and final intervals  by $\bar{v}\Delta t=|Q|$,
one also has that $\bar{v}\ll c$.
When the duration $\Delta t$ of the initial and final time intervals is
decreased,
which needs to be done in order to approach the BR steplike trajectory,
the constant $Q$, while staying finite, must decrease accordingly for a
given test body's maximum speed $v_{\rm max}$.
The average force (3) can now be written as
\begin{eqnarray}
\bar{F}&=& \frac{\rho_c^2V^2}{\tau}\left[
Q\int_{\Delta t}^{\tau-\Delta t}dt_1\,f(t_1)
+\int_0^{\Delta t}dt_1\,Q(t_1)f(t_1)\right.\nonumber \\
&&+\left.\int_{\tau-\Delta t}^{\tau}dt_1\,Q(t_1)f(t_1)\right].
\end{eqnarray}

To simplify their calculations,  BR obtained the average self-force
on the test body assuming for it a strictly  steplike trajectory
$Q_{\rm BR}(t_1)=Q\Theta(t_1)\Theta(\tau-t_1)$,
which is a limit $\Delta t\rightarrow 0$ of our trajectory $Q(t_1)$.
A formal substitution of the BR trajectory $Q_{\rm BR}(t_1)$ in (3) gives
the time-averaged
self-force $\bar{F}_{\rm BR}$, obtained by BR in a different way in
their analysis:
\begin{equation}
\bar{F}_{\rm BR}=\rho_c^2V^2\tau Q\bar{A}^{\rm(I,I)}_{xx},
\end{equation}
where
\begin{eqnarray}
\bar{A}^{\rm(I,I)}_{xx}&=&\frac{1}{\tau^2}\int_0^{\tau}dt_1 f(t_1)\nonumber\\
&=&-\frac{1}{8R^4\kappa}(4+\kappa)(2-\kappa)^2\Theta(2-\kappa)
-\frac{1}{R^4\kappa}, 
\end{eqnarray}
with $\kappa=\tau/R$,
is a BR geometric factor for coinciding spherical space-time
regions, which we here evaluated in closed form. According to Eqs.\ (11)
and (12),  the average BR self-force for a field measurement time
$\tau\ge 2R$ reduces to a force $-\rho_c^2V^2Q/R^3$, which is 
the electrostatic
force of attraction between the test and neutralization bodies when their
centers are displaced by a distance $|Q|\ll R$.
Without the use of a compensating spring, Eq.\ (12) [together with Eq.\ (48)
of BR] leads to a minimum uncertainty
$\Delta\bar{\cal E}_x\sim (\hbar|\bar{A}^{\rm(I,I)}_{xx}|)^{1/2}
\sim (\hbar/\tau V)^{1/2}$ in the measured field component $\bar{\cal E}_x$
for both $\tau\ge 2R$ and $\tau<2R$---which in fact agrees with
the uncertainty (28) of CP, obtained by them for $\tau>2R$ \cite{BM}.

CP contend that the BR use of the steplike trajectory, which
leads to the BR result (11), is incorrect, presumably  as it
implies that the velocity of the test body diverges in the
vicinities of the beginning $t_1=0$ and end $t_1=\tau$ of the measurement
period. However, with our evaluations (10) and (12) of the average self-force
$\bar{F}$ and BR geometric factor $\bar{A}^{\rm(I,I)}_{xx}$,
it is easy to show that the BR self-force
approximates correctly the self-force obtained with a ``physical"
trajectory of a sufficiently steplike character.
Dividing the average self-force (10) by the BR average self-force
(11), we get
\begin{equation}
\frac{\bar{F}}{\bar{F}_{\rm BR}}=
\frac{1}{\bar{A}^{\rm(I,I)}_{xx}\tau^2}\int_{\Delta t}^{\tau-\Delta t}dt_1\,
f(t_1)
+\frac{\Delta\bar{F}_i}{\bar{F}_{\rm BR}}
+\frac{\Delta\bar{F}_f}{\bar{F}_{\rm BR}},
\end{equation}
where the quantities $\Delta\bar{F}_i$  and $\Delta\bar{F}_f$
arise from the time intervals of duration $\Delta t$ at the beginning
$t_1=0$ and end $t_1=\tau$ of the trajectory, respectively.
We find easily an upper bound on
the absolute value of the quantity $\Delta\bar{F}_i$
using the facts that the maximum value of the function $|f(t_1)|$
is $3/R^3$ for  $0\le t_1\le\tau$ [see Eq.\ (9)] and  that
$|Q(t_1)|< v_{\rm max} \Delta t$ in the initial time interval:
\begin{eqnarray}
&&|\Delta \bar{F}_i|=\frac{\rho_c^2V^2}{\tau}\left|\int_0^{\Delta t}dt_1\,
Q(t_1)f(t_1)\right| \nonumber \\
&&\le \frac{\rho_c^2V^2}{\tau}\int_0^{\Delta t}dt_1\,|Q(t_1)|\, |f(t_1)|
<\rho_c^2 V^2v_{\rm max}\frac{3}{R^3}\frac{\Delta t^2}{\tau}.
\end{eqnarray}
We find  in the same way the same upper bound on
the absolute value of the quantity
$\Delta\bar{F}_f=(\rho_c^2V^2/\tau)\int_{\tau-\Delta t}^{\tau}dt_1\,
 Q(t_1)f(t_1)$.
The absolute values of both the ratios
$\Delta\bar{F}_{i,f}/\bar{F}_{\rm BR}$  thus have an upper bound
\begin{eqnarray}
&&\left|\frac{\Delta\bar{F}_{i,f}}{\bar{F}_{\rm BR}}\right|<
\frac{3}{\tau R^3|\bar{A}_{xx}^{\rm (I,I)}|}\,\frac{v_{\rm max}}{\bar{v}}\,
\frac{\Delta t}{\tau}\nonumber \\
&&= \frac{24}{(4+\kappa)(2-\kappa)^2\Theta(2-\kappa)+8}\,
\frac{v_{\rm max}}{\bar{v}}\, \frac{\Delta t}{\tau},
\end{eqnarray}
where we used Eq.\ (11) for $\bar{F}_{\rm BR}$ with $|Q|=\bar{v}\Delta t$
and the closed-form expression (12) for $\bar{A}_{xx}^{\rm (I,I)}$.
As both the speeds $v_{\rm max}$ and $\bar{v}$ may be assumed to be
independent of
$\Delta t$, the upper bound (15) can be made arbitrarily small by letting
$\Delta t$  be sufficiently small, and thus
$\lim_{\Delta t\rightarrow 0}(\Delta\bar{F}_{i,f}/\bar{F}_{\rm BR})=0$.
Using this result, the limit $\Delta t\rightarrow 0$ in Eq.\ (13)
is simply
\begin{equation}
\lim_{\Delta t\rightarrow 0}\frac{\bar{F}}{\bar{F}_{\rm BR}}=1,
\end{equation}
as $\int_0^{\tau}dt_1\,f(t_1) =\tau^2\bar{A}^{\rm(I,I)}_{xx}$. This
means that while both  $\lim_{\Delta t\rightarrow 0}\bar{F}=0$  and
$\lim_{\Delta t\rightarrow 0}\bar{F}_{\rm BR}=0$
[because $\lim_{\Delta t\rightarrow 0}|Q|=
\lim_{\Delta t\rightarrow 0}(\bar{v}\Delta t)=0$],
a ``physical" average self-force $\bar{F}$
obtained with a sufficiently small but finite $\Delta t$ and accordingly
small but finite displacement $|Q|\ll c\Delta t$ is approximated
arbitrarily closely by the BR self-force $\bar{F}_{\rm BR}$ of Eq.\ (11):
\begin{equation}
\bar{F}\approx \bar{F}_{\rm BR}
\;\;\;{\rm when}\;\Delta t\;{\rm is}\;{\rm sufficiently}\;{\rm small.}
\end{equation}

We evaluate also an average force $\bar{F}_Q$, which is the time average
of the force $F_Q(t_2)$ defined by Eq.\ (40) of CP as the component of the
self-force $F(t_2)$ that is directly proportional to the displacement 
$Q(t_2)$. CP show that 
this force is canceled by a force that arises when the neutralizing body is
removed temporarily for the duration of the field measurement.
It is not clear whether a procedure could be devised for
such a removal of the neutralizing body without introducing additional
fields that affect the test body, but we shall leave this point aside.
The average force
$\bar{F}_Q$ can be written as
\begin{equation}
\bar{F}_Q=\frac{\rho_c^2V^2}{\tau}\int_0^{\tau}dt_2\,Q(t_2)g(t_2),
\end{equation}
where
\begin{eqnarray}
g(t_2)&=&-\int\rho(r_1)\,d{\bf r}_1\int\rho(r_2)
\,d{\bf r}_2\frac{\partial^2}{\partial x_1 \partial x_2}
\frac{\Theta(t_2-r)}{r}\nonumber \\
&=&-\frac{1}{2}f(\tau-t_2)-\frac{3}{2R^3}\nonumber \\
&=&\frac{1}{4R^3}(2-\xi)(2-2\xi-\xi^2)\Theta(2-\xi)-\frac{1}{R^3},
\end{eqnarray}
with $\xi=t_2/R$.
Here, the space integration is done simply by using
the result (9) of the space integration in (8) on noting that the 
monopole component of the regularized function
$-\lim_{\epsilon\rightarrow 0}(\partial^2/\partial x_1\partial x_2)
[\Theta(t_2-r)/(r+\epsilon)]$ can be expressed in terms of the function
$\bar{A}^{(1,2)}_{xx\,0}(t_1, r)$ as
\begin{eqnarray}
&&\frac{1}{3}\frac{\delta'(t_2-r)}{r}-\frac{2}{3}\lim_{\epsilon\rightarrow 0}
\frac{\epsilon}{(r+\epsilon)^3}\frac{\Theta(t_2-r)}{r}\nonumber \\
&&=-\frac{1}{2}\bar{A}^{(1,2)}_{xx\,0}(\tau-t_2,r)
-\lim_{\epsilon\rightarrow 0}\frac{\epsilon}{(r+\epsilon)^3}\frac{\Theta
(t_2-r)}{r},
\end{eqnarray}
and that the contribution of the term
$-(2/3)\lim_{\epsilon\rightarrow 0}[\epsilon/(r+\epsilon)^3]
\Theta(\tau-t_1-r)/r$
in Eq.\ (7) to the function $f(t_1)$ of Eq.\ (9) is $-1/R^3$.

According to Eq.\ (19), the function $g(t_2)$ is related in a simple way to 
the function $f(t_1)$,
and thus on the strength of the same argument as that leading to Eq.\ (17),
but using the function $g(t_2)$ instead of the function $f(t_1)$,
it follows that
\begin{eqnarray}
\bar{F}_Q&&\approx \bar{F}_{Q\rm(BR)}=-\frac{1}{2}\bar{F}_{\rm BR}
-\frac{3\rho_c^2V^2Q}{2R^3}\nonumber \\
&&{\rm when}\;\Delta t\;{\rm is}\;{\rm sufficiently}\;{\rm small,}
\end{eqnarray}
where $\bar{F}_{Q\rm(BR)}=(\rho_c^2V^2 Q/\tau)\int_0^{\tau}dt_2\,g(t_2)$
is the average force $\bar{F}_{Q}$ obtained
with the steplike trajectory $Q_{\rm BR}(t_2)=Q\Theta(t_2)\Theta(\tau-t_2)$.
Following CP, we now define an average force
$\bar{F}_{\rm RR}=\bar{F}-\bar{F}_Q$,
which is the time average of what CP call the ``radiation-reaction"
component $F_{\rm RR}(t_2)$ [see Eq.\ (40) of CP] of the self-force $F(t_2)$.
Using Eqs.\ (17) and (21), it is seen easily that
\begin{eqnarray}
\bar{F}_{\rm RR}&&\approx (\bar{F}_{\rm BR}-\bar{F}_{Q\rm(BR)})
=\frac{3}{2}\left(\bar{F}_{\rm BR}+\frac{\rho_c^2V^2Q}{R^3}\right)\nonumber\\
&&\equiv\bar{F}_{\rm RR(BR)}
\;\;\;{\rm when}\;\Delta t\;{\rm is}\;{\rm sufficiently}\;{\rm small.}
\end{eqnarray}
This means that the BR limiting self-force $\bar{F}_{\rm BR}$ has
a ``radiation reaction"  component $\bar{F}_{\rm RR(BR)}$, given
according to Eqs.\ (11), (12) and (22) by
\begin{equation}
\bar{F}_{\rm RR(BR)}=-\frac{3\rho_c^2V^2 Q}{16R^3}(4+\kappa)(2-\kappa)^2
\Theta(2-\kappa),  
\end{equation}
where $\kappa=\tau/R$.
The average ``radiation-reaction" force $\bar{F}_{\rm RR(BR)}$
vanishes only for field measurement times $\tau\ge 2R$.
Now, if the removal of the neutralizing body results in the
cancellation of the force $\bar{F}_Q$, then
the limiting force $\bar{F}_{Q\rm(BR)}$ must be also canceled
and a BR steplike trajectory would result in a net
average self-force $\bar{F}_{\rm RR(BR)}$ of Eq.\ (23),
which, without a compensating spring, would lead to a minimum uncertainty
$\Delta\bar{\cal E}_x\sim (\hbar/\tau R^3)^{1/2}(2-\tau/R)\Theta(2-\tau/R)$
in the measured field component.
The absence of a neutralizing body would result, in the limit of a steplike
trajectory, again
in a time-averaged self-force that is independent of the details of the 
space-time course
of the measurement procedure and, for a field measurement time $\tau<2 R$,
the effect of which would have to be compensated by a BR spring when 
it is desired to measure the field to arbitrary accuracy.
We note here that no use of any neutralizing body, instead of its
possibly problematic temporary removal, would simply subtract 
from the average self-force
$\bar{F}_{\rm BR}$ of Eq.\ (11) the force $-\rho_c^2V^2Q/R^3$ of
electrostatic attraction to the neutralizing body, resulting
in a limiting average self-force that differs only by a factor of $2/3$
from the average self-force $\bar{F}_{\rm RR(BR)}$ of Eq.\ (23) that is obtained
with the temporary removal.

The steplike character of the test-body's trajectory in the
BR analysis is necessitated by the demands on the type of momentum
measurements that have to be performed on the test body at the beginning and
end of the field measurement period $(0,\tau)$. These momentum
measurements are required for the determination of the momentum transfer
along the given direction from the field to the test body,
and are each allowed to have only a duration $\Delta t\ll \tau$.
As BR have shown, the latter requirement is necessary in order to be able
to neglect the radiation reaction on an extended test body during the time
of the momentum measurement.
Thus the momentum measurements are  required to be of the
ideal repeatable type, i.e., for a given precision, of arbitrarily
short duration while at the same time not altering the momentum of the
measured object. BR found in the course of their analysis
a procedure for such repeatable momentum measurements;
a similar procedure was found by Aharonov and Bohm independently some
30 years later \cite{AB,AP}.  A repeatable momentum measurement of accuracy
$\Delta p_x$ and duration $\Delta t$ at the beginning of the field
measurement period $(0,\tau)$ still results in an unpredictable displacement
$Q$ of the test body such that
$|Q|\gtrsim \hbar/\Delta p_x$, occurring within the initial time
interval $(0,\Delta t)$. The requirements that  $|Q|\ll a$ and
$|Q|\ll c\Delta t$ will be satisfied by having the mass of the test body
sufficiently great, and this specification will also guarantee that
the test body can be considered to be essentially at rest
in the interval $(\Delta t, \tau-\Delta t)$ in which it acquires
momentum from the measured field \cite{foot3}.

The test body's trajectory is thus necessarily of a steplike character,
and so the ``radiation-reaction" component $\bar{F}_{\rm RR}$ of its 
average self-force can be approximated by the limiting 
``radiation-reaction" force $\bar{F}_{\rm RR(BR)}$,
which is not affected by the removal of the neutralizing body.
The removal or the absence of the neutralizing body
would not open the possibility of an arbitrarily accurate measurement
of a single field component, averaged over a time $\tau<2R$, without a
compensating spring.

The author gratefully acknowledges correspondence with F. Persico,
whose searching questions helped the author to find correct expressions for
some algebraic results used in the present Comment.


\begin{references}
\bibitem{foot3}
The relation $|Q|\gtrsim \hbar/\Delta p_x$ is not in conflict with the
condition $|Q|\ll c \Delta t$, as $\Delta p_x$ can be arbitrarily great
without impairing the accuracy of the determination
of the measured field component when the test body's charge density
is chosen sufficiently great [see Eq.\ (46) of BR].
\end{references}
\end{document}